\documentclass[aps,prd,twocolumn,preprintnumbers,showpacs,nofootinbib,floatfix,10pt]{revtex4-1}

\pdfoutput=1

\usepackage{textcomp}
\usepackage[english]{babel}
\usepackage{amsmath,amssymb,amsbsy,booktabs}
\usepackage{bm}
\usepackage{color}
\usepackage{array}
\usepackage{amstext}
\usepackage{graphicx}
\usepackage{amsfonts}
\usepackage{bm}
\usepackage{dcolumn}
\usepackage{rotating}
\usepackage{epstopdf}
\usepackage{esint}
\usepackage{url}
\usepackage[colorlinks=true,
linkcolor=blue,
filecolor=blue,
anchorcolor=blue,
urlcolor=blue,
citecolor=blue
]{hyperref}

\usepackage{ulem}

\makeatletter
\newcommand{\thickhline}{%
	\noalign {\ifnum 0=`}\fi \hrule height 1pt
	\futurelet \reserved@a \@xhline
}
\makeatother

\usepackage[utf8]{inputenc}
\usepackage{float}
\usepackage{times}

\allowdisplaybreaks[0]

\begin{document}

	\title {\bf Next-to-tribimaximal mixing against \boldmath${CP}$ violation and baryon asymmetry signs}
	\author{Shao-Ping Li$^{1}$}
    \email{ShowpingLee@mails.ccnu.edu.cn}
    \author{Yuan-Yuan Li$^{1,3}$}
    \email{liyuanyuan@mails.ccnu.edu.cn}
    \author{Xin-Shuai Yan$^{1}$}
    \email{xinshuai@mail.ccnu.edu.cn}
    \author{Xin Zhang$^{2}$}
    \email{xinzhang@hubu.edu.cn (Corresponding author)}
	\affiliation{$^1$Institute of Particle Physics and Key Laboratory of Quark and Lepton Physics~(MOE),\\
		Central China Normal University, Wuhan, Hubei 430079, China\\
        $^2$Faculty of Physics and Electronic Science, Hubei University, Wuhan, Hubei 430062, China\\
        $^3$Basic Department, Information Engineering University,  Zhengzhou, Henan 450000, China}
	
	\begin{abstract}
	Four next-to-tribimaximal (NTBM) mixing patterns  are   widely considered as the feasible candidates for  the  observed leptonic mixing structure.  With the recent measurements from T2K, as well as intensive global fits,  the interval of $\sin \delta_{CP}>0$ for the Dirac $CP$-violating phase is persistently small in the normal ordering  while  $\sin \delta_{CP}<0$ is successively  obtained up  to $3\sigma$ level in the inverted ordering. In this paper, we advocate  the  fitting results of  Dirac $CP$-violating phase as a constraint,  and show that,  it can basically rule out half of the regions allowed by the constraint from three mixing angles only.  In addition, given the  small $\sin \delta_{CP}>0$ interval,  we find that  a unique NTBM pattern can be selected out of  the four candidates  within $3\sigma$ uncertainty of the latest NuFIT 5.1, when the patterns are exposed to a Yukawa texture-independent leptogenesis for the baryon asymmetry of the Universe. For the surviving pattern, we find an interesting correlation between the Dirac $CP$-violating phase and the octant of atmospheric angle $\theta_{23}$, where  $\theta_{23}>45^\circ$ can be predicted  once $5\pi/6<\delta_{CP}<\pi$  in the normal ordering can be  confirmed in the future measurements.

	\end{abstract}

	
	\maketitle

\section{Introduction}
Neutrino oscillation experiments have undoubtedly unveiled nonzero neutrino masses and  nontrivial  leptonic mixing. Thus far, the intention to explain neutrino masses and mixing, or the so-called flavor puzzle,  has triggered a great deal of theoretical investigations beyond the standard model. The flavor symmetry-induced neutrino mixing was  prevailing in the first decade of the twenty-first century (we refer to the recent reviews~\cite{Xing:2019vks,Feruglio:2019ybq} in which early studies can be found). In particular, the so-called tribimaximal (TBM) mixing~\cite{Harrison:2002er},

\begin{align}\label{TBM}
	V_{\text{TBM}} =\left(
	\begin{array}{ccc}
		\sqrt{\frac{2}{3}}\,\,  &  \frac{1}{\sqrt{3}}\,\,   &                   0 \\[0.2cm]
		-\frac{1}{\sqrt{6}}\,\,  & 	\frac{1}{\sqrt{3}}\,\,   &  \frac{1}{\sqrt{2}} \\[0.2cm]
		-\frac{1}{\sqrt{6}}\,\,  & 	\frac{1}{\sqrt{3}}\,\,   & -\frac{1}{\sqrt{2}} \\
	\end{array}
	\right),
\end{align}
was once considered to be the promising candidate for explaining the observed Pontecorvo-Maki-Nakawaga-Sakata (PMNS) matrix, which in the standard parametrization is given by~\cite{Zyla:2020zbs}

\begin{widetext}
	
\begin{align}
	U_{P}=\left(
	\begin{array}{ccc}
		c_{{12}}c_{{13}} \,  &  s_{{12}} c_{{13}} \,   &   e^{-i \delta_{CP}}   s_{{13}} \\[0.2cm]
	-s_{{12}}c_{{23}}-e^{i \delta_{CP}}c_{{12}} s_{{13}} s_{_{23}}   \, &    	c_{{12}}c_{{23}}-e^{i \delta_{CP}}s_{{12}} s_{{13}} s_{{23}}\,&   c_{{13}} s_{{23}} \\[0.2cm]
	s_{{12}}s_{{23}}-e^{i \delta_{CP}}c_{{12}} s_{{13}} c_{{23}}  \, & 	-c_{{12}}s_{{23}}-e^{i \delta_{CP}}s_{{12}} s_{{13}} c_{{23}}\,  & c_{{13}}c_{{23}}\\
	\end{array}
	\right).
\end{align}

\end{widetext}

Here, $s_{ij}\equiv \sin\theta_{ij}, c_{ij}\equiv \cos\theta_{ij}$, $\delta_{CP}$ is the Dirac $CP$-violating phase, and we have neglected the possible Majorana phases as they do not appear in neutrino oscillations.  After the discovery of nonzero $\theta_{13}$~\cite{T2K:2011ypd,MINOS:2011amj,DoubleChooz:2011ymz,DayaBay:2012fng,RENO:2012mkc}, the TBM mixing that predicts $\theta_{13}=0$ is ruled out in the neutrino mixing sector  where the charged-lepton Yukawa matrix is diagonal. Later on,  TBM variants were extensively investigated to accommodate   $\theta_{13}\neq 0$ and reproduce the PMNS matrix. Among them, the minimal corrections to TBM mixing, or the next-to-TBM (NTBM)~\cite{Albright:2008rp,He:2011gb,Rodejohann:2012cf}, which are characterized by a two-parameter family, have brought us certain predictions and correlations among the mixing angles $\theta_{12,13,23}$ and phase $\delta_{CP}$, and have been shown to be able to reproduce the PMNS matrix in model-independent analyses (see, e.g., Refs.~\cite{Chao:2011sp,Rodejohann:2011uz,He:2011gb,Araki:2011wn,Kang:2014mka,Garg:2017mjk,Garg:2018jsg,Chen:2018eou} and references therein).

In recent years, experiments from T2K~\cite{T2K:2018rhz,T2K:2019bcf} and the global fits in the years 2018-2021~\cite{Capozzi:2018ubv,Esteban:2018azc,deSalas:2020pgw,Esteban:2020cvm,Capozzi:2021fjo}  have persistently indicated a small parameter region for $\sin\delta_{CP}>0$ in the normal ordering (NO) while $\sin\delta_{CP}<0$ is successively obtained in the inverted ordering (IO)  within $3\sigma$ level of  uncertainty. This is further confirmed by the latest NuFIT 5.1~\cite{nufit5.1:2021,Gonzalez-Garcia:2021dve}. Nevertheless, the true value of $\delta_{CP}$ is still unknown, and this ambiguity partly allows earlier analyses of TBM variants to take $\delta_{CP}$  as a \textit{prediction} rather than a \textit{constraint}~\cite{Kang:2014mka,Garg:2018jsg,Zhao:2018vxy,Garces:2018nar,Garg:2018rfz,Everett:2019idp,Rock:2021xis}. Given the persistent results, in this paper we will instead take $\delta_{CP}$ as a  constraint, and show in a model-independent way, that the viable parameter space of the  NTBM patterns significantly shrinks.

While imposing the  constraint of  leptonic $CP$ violation   would cut down nearly half of the parameter space, none of  the four NTBM patterns can be uniquely selected out,   which  is also a common situation in various model-independent analyses. To pin down a unique  NTBM,  we find that, only one pattern survives if the sign of leptonic $CP$ violation matches $\sin\delta_{CP}>0$. Noticeably, the sign of leptonic $CP$ violation  can  be  closely  related  to the sign of baryon asymmetry in the Universe (BAU) via leptogenesis~\cite{Fukugita:1986hr} .  In fact, it has been shown in a   Yukawa texture-independent Dirac leptogenesis~\cite{Li:2021tlv}, that  a positive BAU requires $\sin\delta_{CP}>0$.

The motivation for applying Yukawa texture-independent   leptogenesis as a selection criterion goes as follows. Lepton Yukawa matrices are the building blocks for the generic leptogenesis mechanism (see e.g., the review~\cite{Davidson:2008bu}), and meanwhile account for the observed lepton masses and PMNS matrix. For Yukawa texture-independent   leptogenesis,  there exists a direct link between the BAU and PMNS matrix such that the resulting baryon asymmetry is directly attributed to   the   PMNS structure. Then, for a broad class of theoretical PMNS candidates, such as the NTBM patterns,  the BAU selection criterion can potentially provide an additional constraint on the PMNS structures.

	A direct BAU-PMNS connection is,     nevertheless, still  under extensive  investigations at present day.  A generic way is to invoke some  assumptions or parametrizations  of  the lepton  Yukawa structures, as  widely considered in Refs.~\cite{Buchmuller:2000as,Joshipura:2001ui,Davidson:2001zk,Moffat:2018smo,Xing:2020ghj,Xing:2020erm,Rahat:2020mio,Gu:2006dc,Gu:2007mi,Wang:2016lve,Li:2020ner}. However, the corresponding BAU selection criterion would then crucially depend on the assumptions and parametrizations,  or even be irrelevant to the Dirac $CP$-violating phase~\cite{Branco:2001pq,Rebelo:2002wj,Branco:2011zb}.  Therefore, to build a robust BAU selection criterion, the prediction of baryon asymmetry should not depend on  nontrivial  Yukawa textures. A purely thermal Dirac leptogenesis considered in Ref.~\cite{Li:2021tlv} is a sound candidate in this respect. The corresponding mechanism formulates  the baryon asymmetry in terms of charged-lepton and Dirac neutrino masses and the  PMNS matrix, such that the sign of  baryon asymmetry  uniquely depends on the sign of  $\sin\delta_{CP}$. Besides, the mechanism is  free from any underlying flavor theory that can explain  the lepton mass spectrum and PMNS matrix. In this respect, the Yukawa texture-independent leptogenesis can also play a significant role in guiding the flavor model buildings, especially when the BAU criterion can uniquely select a theoretical PMNS candidate.

 It will be shown that, the Yukawa texture-independent Dirac leptogenesis as a BAU   criterion can finally select a unique NTBM with a narrow parameter space in the NO spectrum, and disfavors all  the patterns in the IO spectrum. Certainly, the approach presented here is not only  valid for  the NTBM patterns, but  also applicable  for other two-parameter family candidates, such as the generalization to the bimaximal mixing~\cite{Barger:1998ta}.   Besides, the suggestion of taking the sign of leptonic $CP$ violation as a constraint should be taken seriously if the upcoming measurements and global fits further confirm previous results, and it   would become more significant, when being assisted with  some Yukawa  texture-independent leptogenesis, to phase out most of, or even all of,  the theoretical PMNS candidates.

The BAU criterion can help to reveal the three unknowns in current neutrino oscillations, i.e., NO versus IO, the $CP$ violation, and  $\theta_{23}<45^\circ$ versus $\theta_{23}>45^\circ$~\cite{Capozzi:2018ubv}. In addition to favoring  an NO spectrum and $\sin\delta_{CP}>0$,   the unique  NTBM pattern selected from  the BAU criterion    further exhibits an intriguing correlation between the Dirac $CP$-violating phase and the  octant of atmospheric angle $\theta_{23}$. It will be shown that, if  $5\pi/6<\delta_{CP}<\pi$  in the NO spectrum can be  confirmed in the future measurements, $\theta_{23}>45^\circ$ will be predicted.

After this Introduction, we firstly review in Sec.~\ref{2pTB} the four NTBM patterns and  formulate the angles and phase entirely in terms of the two free parameters. In Sec.~\ref{CP&cons}, we impose, in a model-independent way,  the  constraints of mixing angles and the Dirac $CP$-violating phase on the four NTBM patterns, and  then expose the survived parameter space to the Yukawa texture-independent leptogenesis in Sec.~\ref{DiracBAU}. The correlation between $\delta_{CP}$ and $\theta_{23}$ will be analyzed in Sec.~\ref{Correlaions}.
Finally we present our conclusions in Sec.~\ref{cons}.

\section{Two-parameter family of  NTBM patterns} \label{2pTB}
There are six NTBM patterns defined by multiplying Eq.~\eqref{TBM} by a unitary rotation matrix on the left or right. However,   the  patterns: $U_P=V_{\text{TBM}}R_{12}$ and $U_P=R_{23}V_{\text{TBM}}$ are already excluded since they predict $\theta_{13}=0$.  The remaining four patterns are denoted as
\begin{align}\label{TBM variants}
	&\text{TBM}_1:& U_P=V_{\text{TBM}}R_{23},~~&\text{TBM}_2:&U_P=V_{\text{TBM}}R_{13},
\nonumber	\\
&\text{TBM}^2:&U_P=R_{13}V_{\text{TBM}},~~ &\text{TBM}^3:& U_P=R_{12}V_{\text{TBM}}.
\end{align}
Here, the rotation matrices are defined as
\begin{align}
	R_{12}&=\left(
	\begin{array}{ccc}
		c_\alpha & s_\alpha e^{i\varphi} & 0\\[0.1cm]
		-s_\alpha e^{-i\varphi }  & c_\alpha  & 0\\[0.1cm]
		0  & 0 & 1\\
	\end{array}
	\right),
	\nonumber \\
	R_{13}&=\left(
	\begin{array}{ccc}
		c_\alpha  & 0     &  s_\alpha e^{i\varphi } \\[0.1cm]
		0  & 1 	& 0 \\[0.1cm]
		-s_\alpha e^{-i\varphi} &   0   & 	c_\alpha\\
	\end{array}
	\right),
		\nonumber \\
		R_{23}&=\left(
	\begin{array}{ccc}
		1 &                   0           &                 0\\[0.1cm]
		0 &  c_\alpha                    &  s_\alpha e^{i\varphi}\\[0.1cm]
		0 & -s_\alpha e^{-i\varphi }    &  c_\alpha\\
	\end{array}
	\right),
\end{align}
where $c_\alpha\equiv \cos\alpha$ and  $s_\alpha\equiv \sin\alpha$, with  $0\leqslant\alpha\leqslant\pi$ and $0\leqslant\varphi<2\pi$. The free parameters $\alpha, \varphi$ characterize the two-parameter family of NTBM patterns. Note that,  $U_P$ defined above has taken into account both the charged-lepton and  neutrino  mixing. In this respect, the pattern, e.g., $\text{TBM}^2$ may be interpreted as TBM neutrino mixing corrected by charged-lepton $e$-$\tau$ mixing~\cite{Xing:2019vks,Feruglio:2019ybq}.
In the following, we will give the formulas of mixing angles and $CP$-violating phase in terms of $\alpha, \varphi$ for the remaining four NTBM patterns, respectively.

  \begin{figure*}[t]
	\centering
	\includegraphics[width=0.45\linewidth]{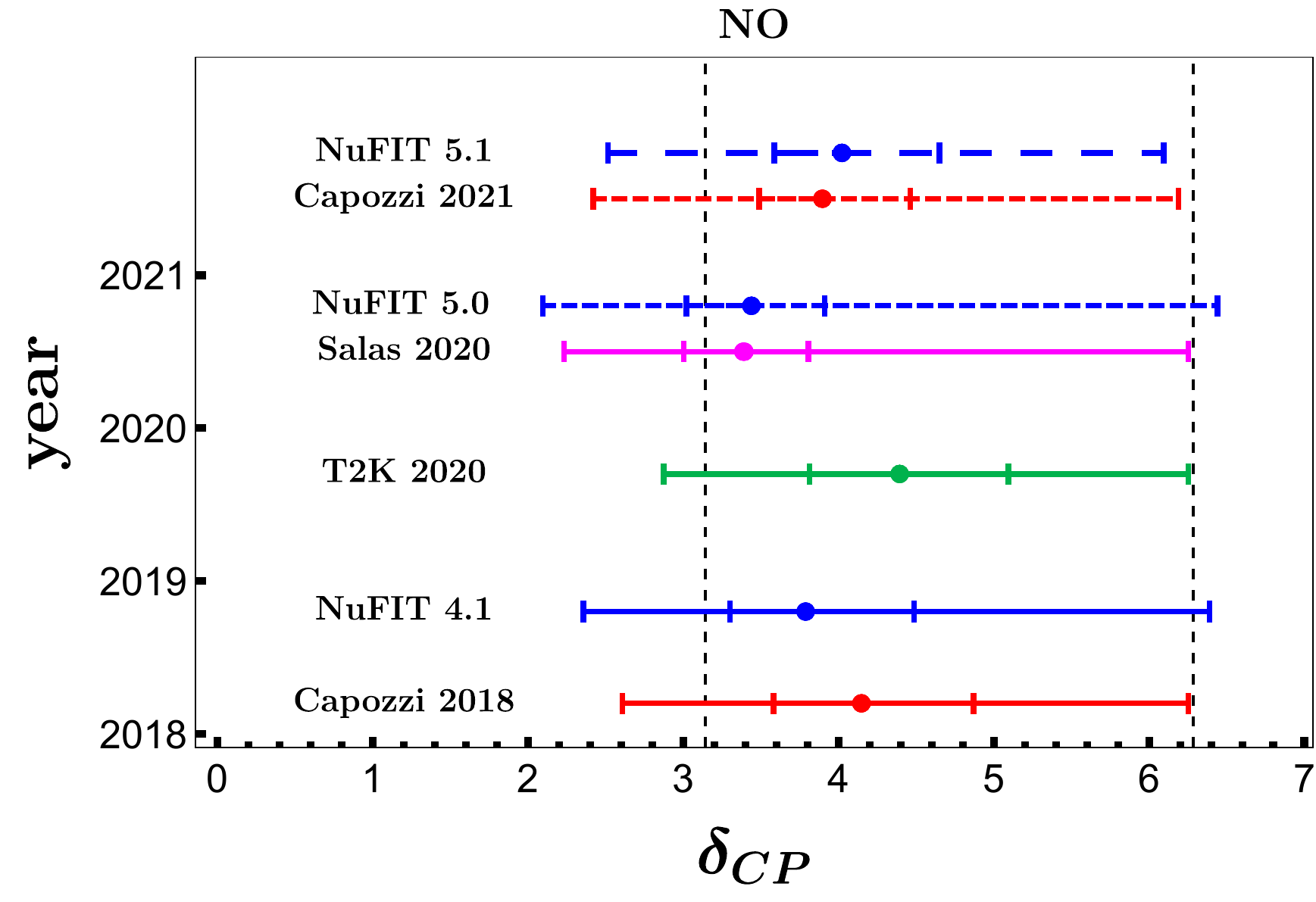}\qquad
	\includegraphics[width=0.45\linewidth]{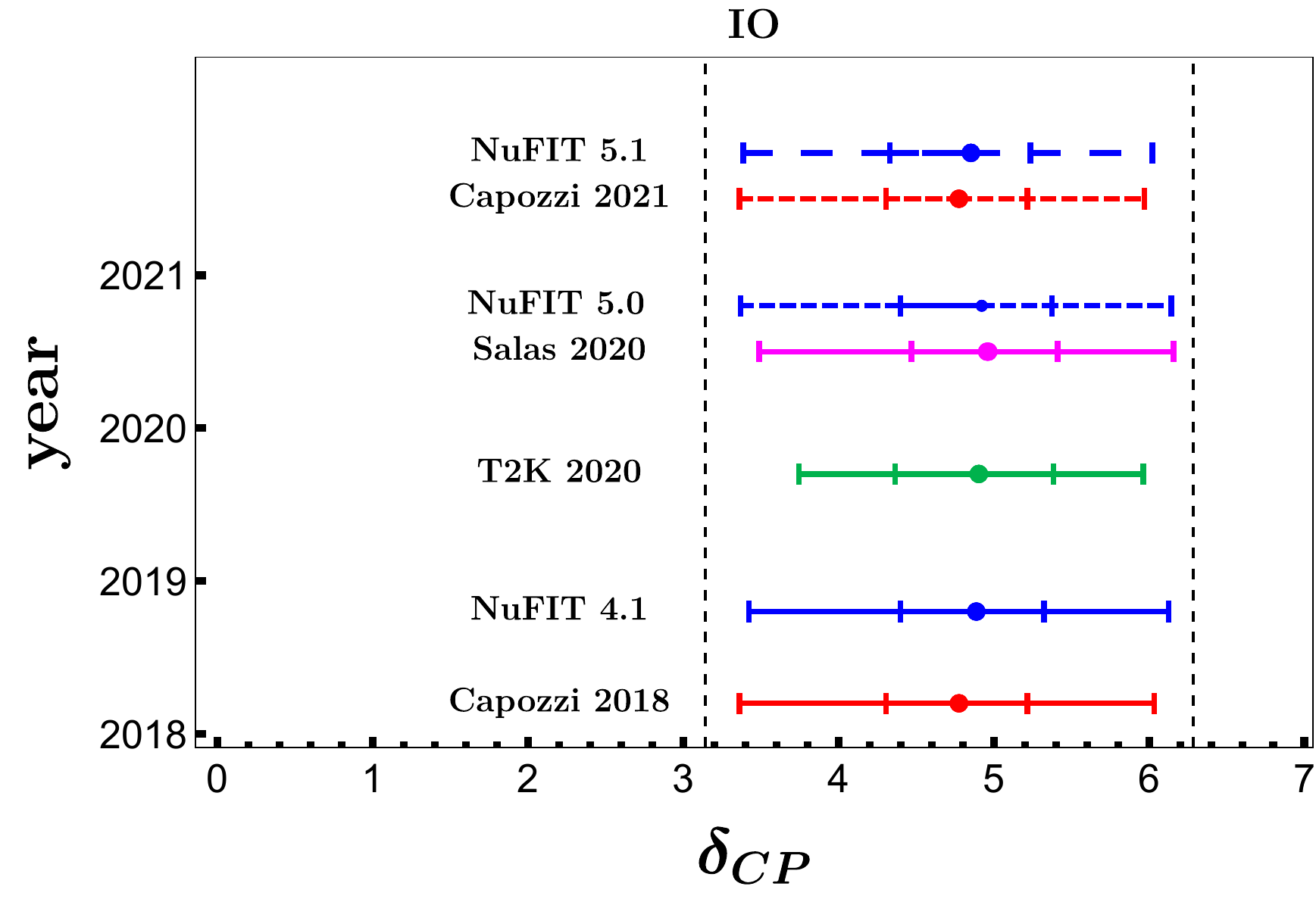}\\
	\caption{\label{CPtrend} The results of Dirac $CP$-violating phase   $\delta_{CP}$ obtained in the years 2018-2021 within $1\sigma$ and $3\sigma$ uncertainties~\cite{Capozzi:2018ubv,Esteban:2018azc,deSalas:2020pgw,Esteban:2020cvm,Capozzi:2021fjo,nufit5.1:2021,Gonzalez-Garcia:2021dve}.  The two vertical lines represent  $\delta_{CP}=\pi, 2\pi$, respectively. }
\end{figure*}

\subsection{$\text{TBM}_1$}
The desired formulas can be derived by using the two simplest rephasing invariant observables, namely, the moduli of PMNS matrix, $|U_P|$,  and the Jarlskog invariant, $\mathcal{J}_{CP}$~\cite{Jarlskog:1985ht}. Explicitly, the mixing angles can be directly obtained by observing the module ratios of PMNS matrix elements via
\begin{align}\label{anglefun-TBM_1}
	\tan \theta_{23}&=\frac{|U_{P,23}|}{|U_{P,33}|}=\left|\frac{3\sqrt{2}c_\alpha+2
		\sqrt{3}e^{i\varphi}s_\alpha}{3\sqrt{2}c_\alpha-2
		\sqrt{3}e^{i\varphi}s_\alpha}\right|,
	\nonumber \\[0.5mm]
	\tan\theta_{12}&=\frac{|U_{P,12}|}{|U_{P,11}|}=\frac{|c_\alpha|}{\sqrt{2}},~
\sin\theta_{13}=|U_{P,13}|=\frac{|s_\alpha|}{\sqrt{3}}.
\end{align}
For the Dirac $CP$-violating phase, we can equate the  $\mathcal{J}_{CP}$ of $U_P$, which is given by~\cite{Zyla:2020zbs}
\begin{align}\label{standardJCP}
	\mathcal{J}_{CP}=\frac{1}{8}\cos\theta_{13}\sin(2\theta_{12})\sin(2\theta_{23}) \sin(2\theta_{13})\sin\delta_{CP},
\end{align}
with the $\mathcal{J}_{CP}$ of $\text{TBM}_1$:
	$\mathcal{J}_{CP}=-s_{2\alpha}s_\varphi/6\sqrt{6}$,
leading to
\begin{align}\label{sinCP-TBM_1}
	\sin\delta_{CP}=-\frac{\text{sgn}(s_{2\alpha})s_\varphi (c_{2\alpha}+5)}{\left[ (c_{2\alpha}+5)^2-(2\sqrt{6}s_{2\alpha} c_\varphi)^2 \right]^{1/2}},
\end{align}
where  $\text{sgn}(s_{2\alpha})$ is the sign function of $s_{2\alpha}\equiv \sin(2\alpha)$.
It should be pointed out that, in deriving Eq.~\eqref{sinCP-TBM_1}, we have simply replaced the mixing angles by using Eq.~\eqref{anglefun-TBM_1} in the first quadrant, e.g., $\theta_{13}=\sin^{-1}(|s_\alpha|/\sqrt{3})$,
since the solution  in the second  quadrant $\theta_{13}=\pi-\sin^{-1}(|s_\alpha|/\sqrt{3})$ leads to the same result in Eq.~\eqref{sinCP-TBM_1} due to the periodicity of $\theta_{13}$ function in Eq.~\eqref{standardJCP}, i.e.,
$\cos\theta_{13}\sin(2\theta_{13})=\cos(\pi-\theta_{13})\sin[2(\pi-\theta_{13})]$. Similar conclusions also hold for $\theta_{12}$ and $\theta_{23}$ functions.

\subsection{$\text{TBM}_2$}
Similar to the above logic, it is straightforward to obtain the formulas for the $\text{TBM}_2$ pattern. The angles are given by
\begin{align}\label{anglefun-TBM_2}
	\tan \theta_{23}&=\left|\frac{\sqrt{3}e^{i\varphi}s_\alpha-3
	 c_\alpha}{3 c_\alpha+
		\sqrt{3}e^{i\varphi}s_\alpha }\right|,
	\nonumber \\
	\tan\theta_{12}&=\frac{1}{\sqrt{2}}\frac{1}{|c_\alpha|},~
	\sin\theta_{13}=\sqrt{\frac{2}{3}}|s_\alpha|,~
\end{align}
  and the $CP$-violating phase can be  simplified as
\begin{align}\label{sinCP-TBM_2}
	\sin\delta_{CP}=- \frac{\text{sgn}(s_{2\alpha})(c_{2\alpha}+2)s_\varphi}{[(2c_\alpha^2+1)^2-3s_{2\alpha}^2c_\varphi^2]^{1/2}},
\end{align}
which is derived by using the $\mathcal{J}_{CP}$ of $\text{TBM}_2$:
$\mathcal{J}_{CP}=-s_{2\alpha}s_\varphi/6\sqrt{3}$.

\subsection{$\text{TBM}^2$}
For the $\text{TBM}^2$ pattern, it can be shown that,
\begin{align}\label{anglefun-TBM^2}
	\tan \theta_{23}=\frac{1}{|c_\alpha|},
	\frac{\tan\theta_{12}}{\sqrt{2}}=\left|\frac{ c_\alpha+e^{i\varphi}
		s_\alpha }{ 2 c_\alpha-
		e^{i\varphi}s_\alpha}\right|,
	\sin\theta_{13}=\frac{|s_\alpha|}{\sqrt{2}},
\end{align}
for the mixing angles. Equating the $\mathcal{J}_{CP}$ in $\text{TBM}^2$:
$\mathcal{J}_{CP}=s_{2\alpha}s_\varphi/12$ with Eq.~\eqref{standardJCP} gives
\begin{align}\label{phasefun-TBM^2}
	\sin\delta_{CP}= \frac{\text{sgn}(s_{2\alpha}) (c_{2\alpha}+3) s_\varphi}{2[(-2 s_{2\alpha} c_\varphi +3c_\alpha^2+1)(1+s_{2\alpha}c_\varphi
		)]^{1/2}}.
\end{align}

\begin{figure*}[th]
	\centering
	\includegraphics[width=0.41\linewidth]{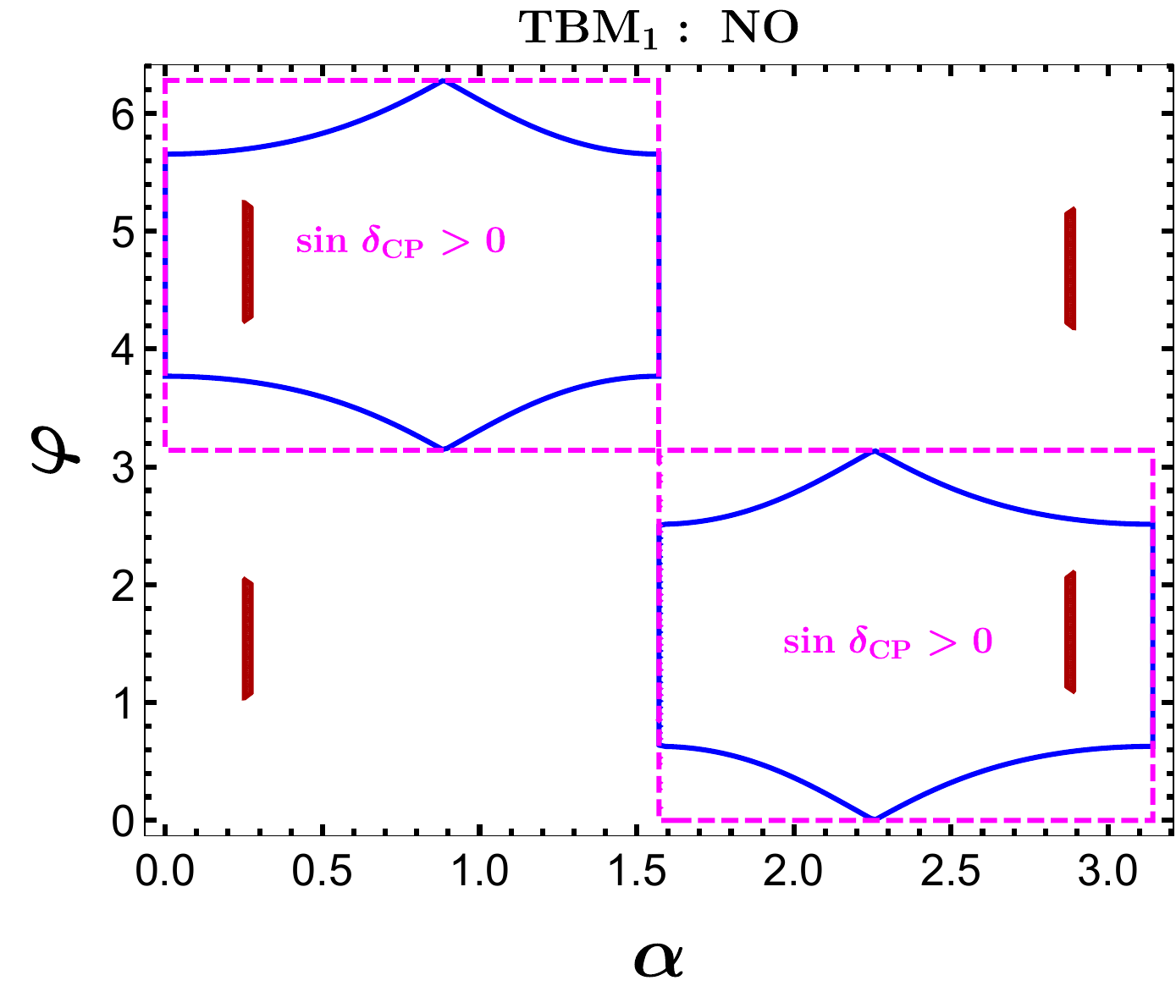}\qquad
	\includegraphics[width=0.41\linewidth]{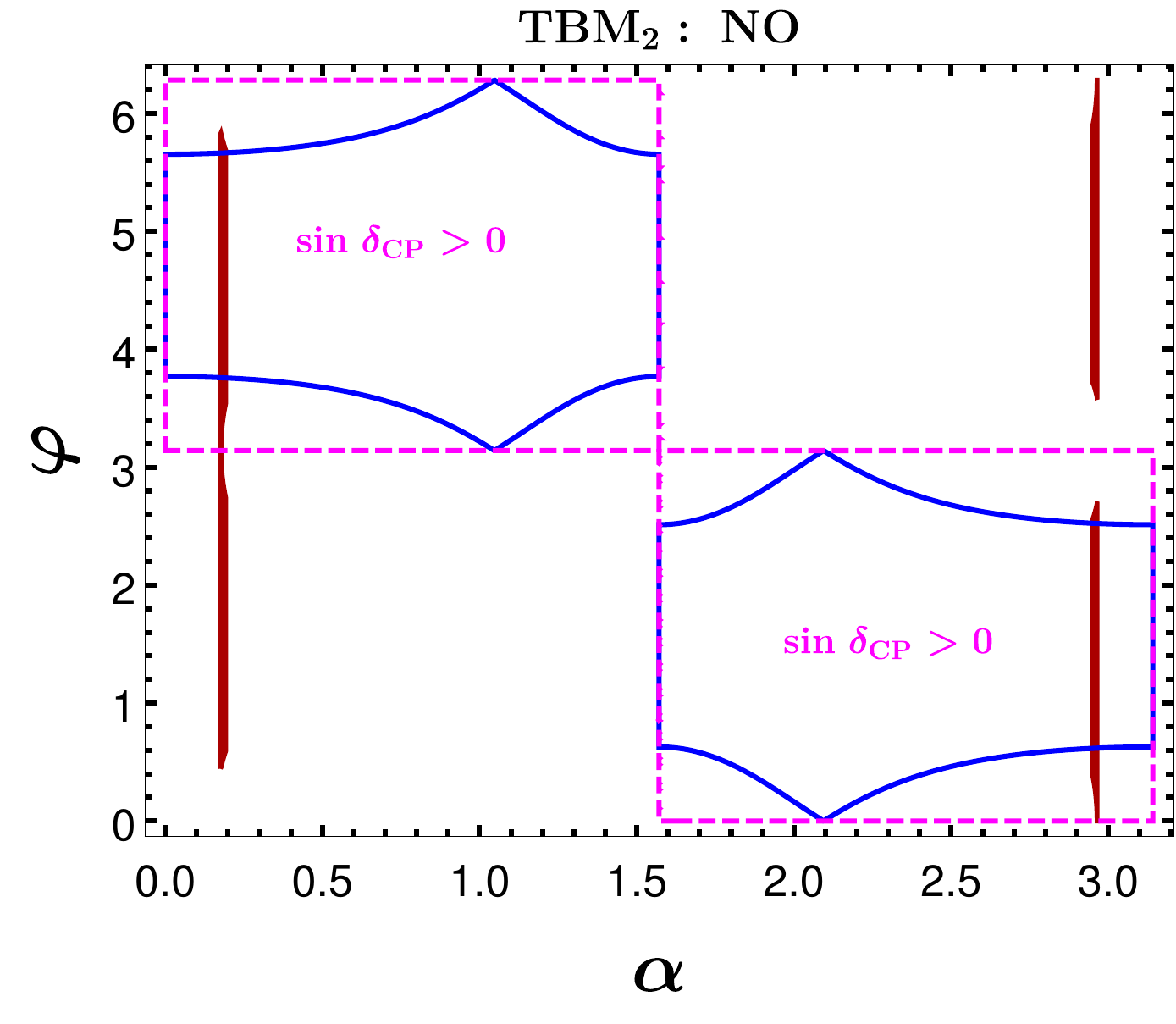}\\
	\includegraphics[width=0.41\linewidth]{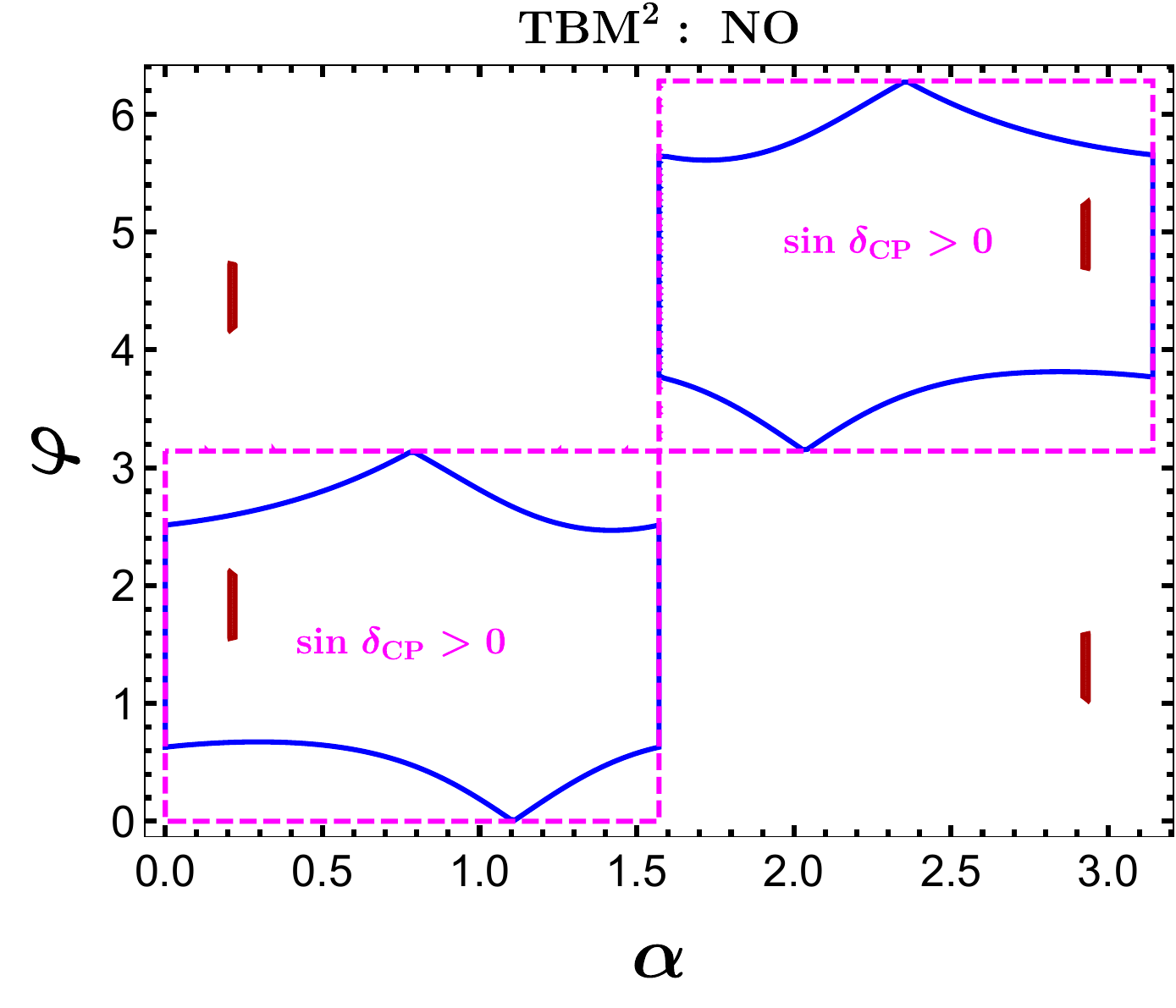}\qquad
	\includegraphics[width=0.41\linewidth]{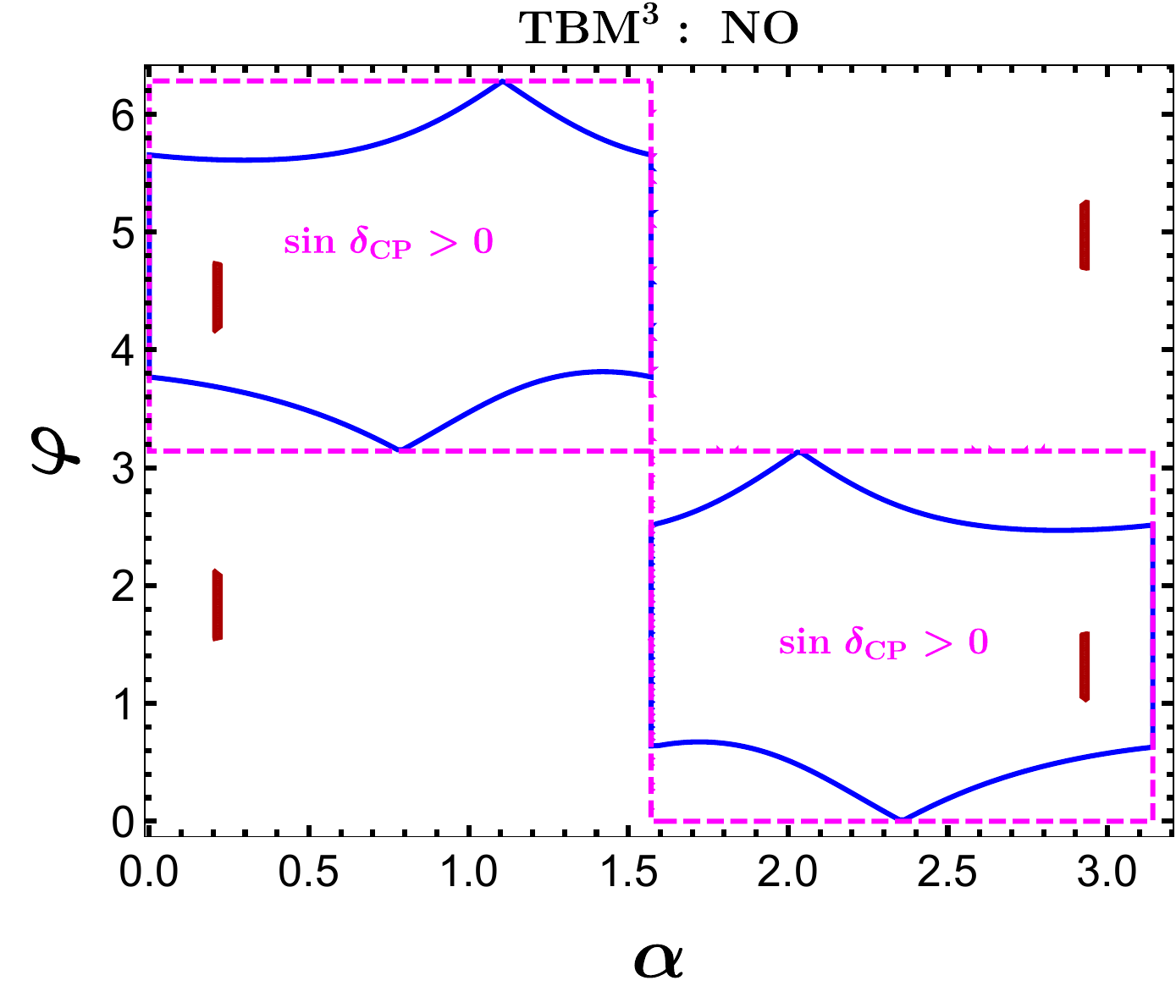}\\
	\caption{\label{NOCPcon} Constraints on the two free parameters $\alpha, \varphi$ in the NO   spectrum. The $3\sigma$-allowed mixing angles are shown in narrow red bands and the regions within the blue contours are excluded by  $\sin\delta_{CP}$ at $3\sigma$ level from NuFIT 5.1~\cite{nufit5.1:2021,Gonzalez-Garcia:2021dve}. For clearness, the regions of $\sin\delta_{CP}>0$ are enclosed by magenta lines.}
\end{figure*}

\subsection{$\text{TBM}^3$}
Finally, for the $\text{TBM}^3$ pattern, the functions for mixing angles are given by
\begin{align}\label{anglefun-TBM^3}
	\tan \theta_{23}= |c_\alpha|,
	\frac{\tan\theta_{12}}{\sqrt{2}}= \left|\frac{c_\alpha+e^{i\varphi}
		s_\alpha}{2 c_\alpha-
		e^{i\varphi}s_\alpha}\right|,
	\sin\theta_{13}=\frac{|s_\alpha|}{\sqrt{2}},~
\end{align}
where $\theta_{12}$ and $\theta_{13}$ have the same expressions as in $\text{TBM}^2$.
The function for phase is
\begin{align}
	\sin\delta_{CP}= -\frac{\text{sgn}(s_{2\alpha}) (c_{2\alpha}+3) s_\varphi}{2[(-2 s_{2\alpha} c_\varphi +3c_\alpha^2+1)(1+s_{2\alpha}c_\varphi
		)]^{1/2}},
\end{align}
which has a different sign  from Eq.~\eqref{phasefun-TBM^2}, as the $\mathcal{J}_{CP}$ in the ${\text{TBM}^3}$ pattern: $\mathcal{J}_{CP}=-s_{2\alpha}s_\varphi/12$ is opposite to that in ${\text{TBM}^2}$.

With these formulas, we can use the two free parameters $\alpha, \varphi$ to fully describe the mixing angles and Dirac $CP$-violating phase extracted from neutrino oscillation experiments.

\begin{figure*}[t]
	\centering
	\includegraphics[width=0.43\linewidth]{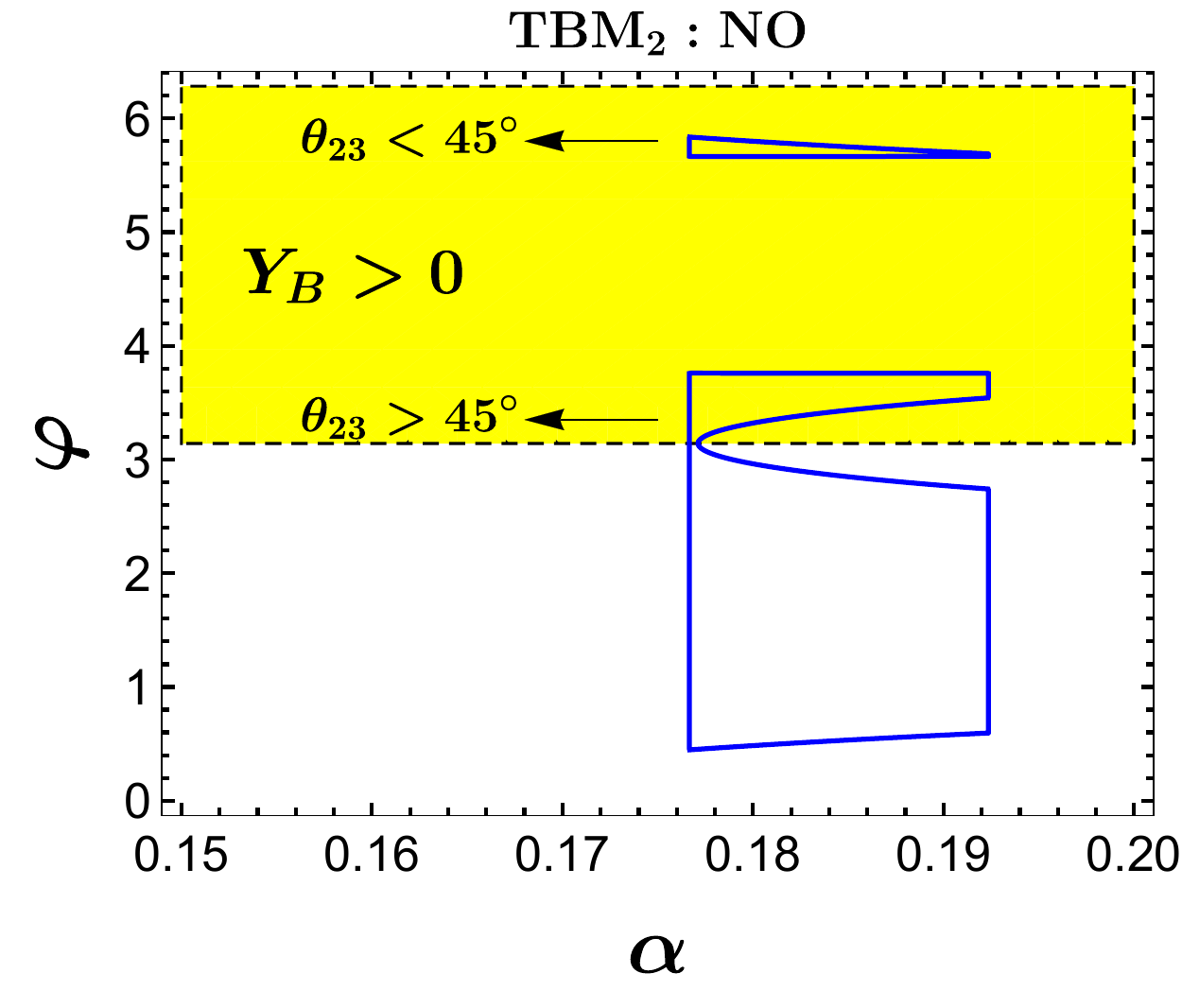}\qquad
	\includegraphics[width=0.43\linewidth]{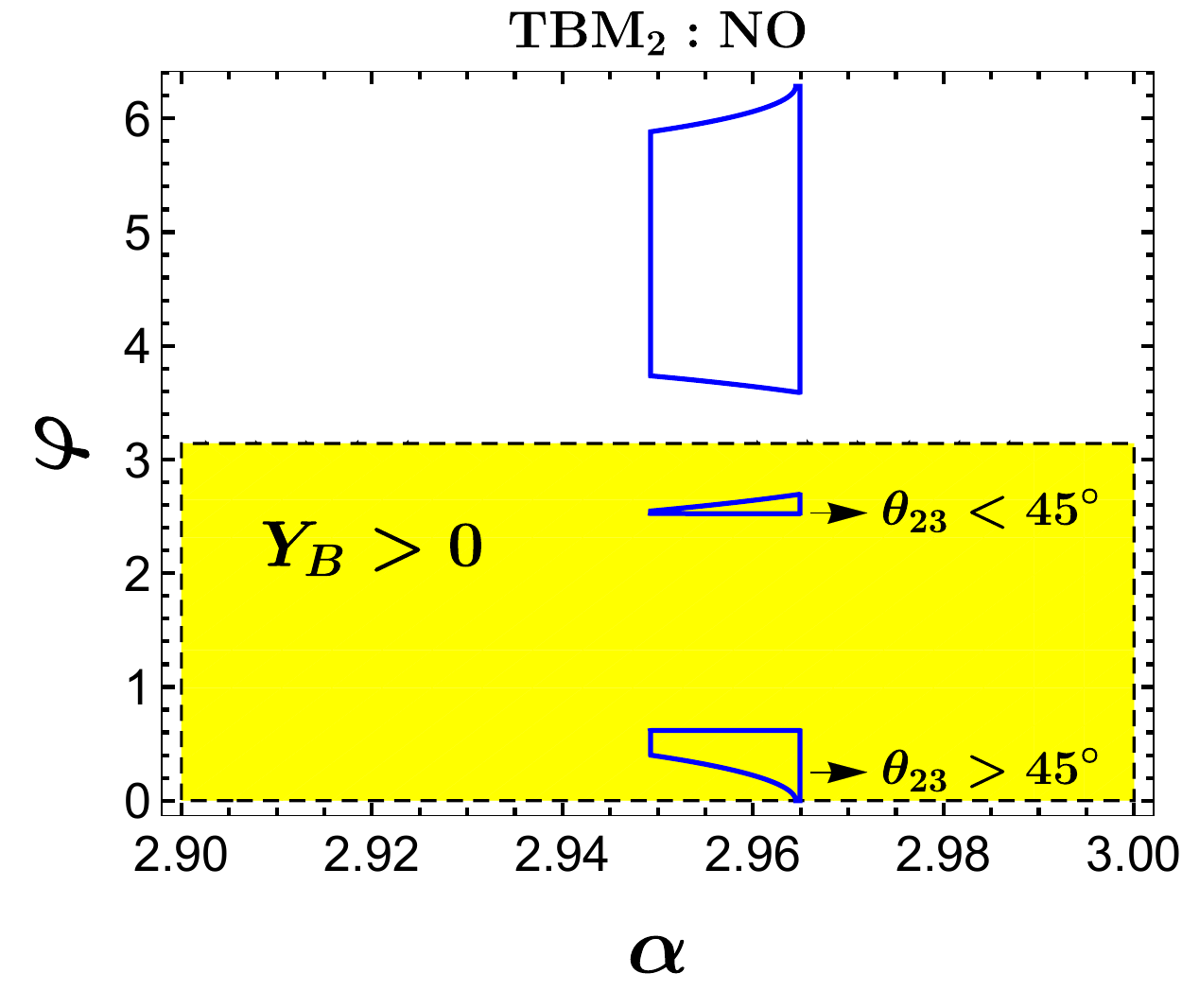}
	\caption{\label{BAUfit} The  $\text{TBM}_{2}$ pattern survives from  the BAU selection criterion. Regions for $Y_{B}>0$ are shown in yellow band, and the survived parameter space from  constraints of mixing angles and Dirac $CP$-violating phase is shown within the blue contour. For visibility,  we have shown the results  in  two panels  with  $\alpha$ in the first and second quadrants, respectively. The asymmetric contours in the $Y_{B}>0$ region correspond respectively to $\theta_{23}<45^\circ$ and $\theta_{23}>45^\circ$ octant.}
\end{figure*}
\section{Leptonic $CP$ violation as a constraint}\label{CP&cons}
Concerning the Dirac $CP$-violating phase, it has been shown over the past few years that, the global fits for  $\delta_{CP}$ have exhibited a rather persistent tendency in a sense that,   the region of $\sin \delta_{CP}>0$ is small in the NO spectrum while   $\sin \delta_{CP}<0$ is stably inferred in the IO spectrum.   We depict in Fig.~\ref{CPtrend} the $\delta_{CP}$ results obtained from T2K measurement~\cite{T2K:2019bcf} and the global fits obtained from 2018 to present day~\cite{Capozzi:2018ubv,Esteban:2018azc,deSalas:2020pgw,Esteban:2020cvm,Capozzi:2021fjo}. It is seen that, for the NO  spectrum,  only a small interval allows $\sin\delta_{CP}>0$ within the $3\sigma$ uncertainty, while for the IO, $\sin\delta_{CP}<0$ is persistently obtained  up to $3\sigma$ uncertainty. Taking these results  as   constraints, we would like to show how $\sin\delta_{CP}$ restricts the parameter space of the NTBM patterns. To this end, we apply the formulas of angles and phase obtained in Sec.~\ref{2pTB} to reproduce the oscillation data at $3\sigma$ level. For concreteness, we apply the latest NuFIT 5.1 result~\cite{nufit5.1:2021,Gonzalez-Garcia:2021dve}, while a comparison between different results will be made whenever necessary.

 We show in Fig.~\ref{NOCPcon} the  survived parameter space for each NTBM pattern under the $3\sigma$-allowed mixing angles (red bands) and $CP$-violating phase in the NO spectrum.  It can be seen that,  except for $\text{TBM}_2$, the constraint from $CP$-violating phase (regions within blue contour) excludes half of the survived parameter space for other three patterns.
 For the IO spectrum, on the other hand, the mixing-angle data are similar to the NO spectrum, and hence the red bands basically coincide. However, since $\sin\delta_{CP}<0$ is maintained within $3\sigma$ uncertainty in the IO spectrum, only the red bands outside the magenta lines are allowed.

 The conclusion thus far is that, with the global results obtained in recent years, the allowed parameter space for the four  NTBM patterns is already quite small, even though the four patterns are still viable if  $3\sigma$ uncertainties are adopted.  The constraint  from $CP$-violating phase is strong and can exclude nearly half of the  already allowed  parameter space where   only three mixing angles are considered.   The survived regions are quite small, especially predicting a rotation angle $0.1< \alpha <0.3$ (as well as its periodic correspondence $0.1< \pi-\alpha <0.3$)   for all the four NTBM patterns.

Due to the currently observed correlation between the mass hierarchy and the sign of $\sin\delta_{CP}$, it can be seen from Fig.~\ref{NOCPcon} that, if the IO spectrum  is confirmed in the future and the data of mixing angles do not change significantly, then none of the four NTBM patterns can be  uniquely selected out. However,   if the NO spectrum is confirmed,  then only the $\text{TBM}_2$ pattern will survive provided that $\sin\delta_{CP}>0$ in the NO spectrum still exists. In fact,  $\sin\delta_{CP}>0$ can be supported in some leptogenesis scenarios, and particularly in a lepton Yukawa texture-independent leptogenesis.

\section {  Yukawa texture-independent BAU As A criterion} \label{DiracBAU}
Baryogenesis through leptogenesis~\cite{Fukugita:1986hr} is the mechanism to
account for the observed BAU today~\cite{Planck:2018vyg}:
\begin{align}\label{YB}
	Y_{B}^{\rm exp}\equiv \frac{n_B-n_{\bar B}}{n_\gamma}\approx 8.75\times 10^{-11}>0,
\end{align}
where $n_{B} (n_{\bar B})$ is the baryon (antibaryon) number density, which is normalized to photon density $n_\gamma$. The generic building block to generate baryon asymmetry  in leptogenesis is the lepton Yukawa matrices.  It is known that, the lepton Yukawa textures are  responsible for the nontrivial PMNS mixing observed in neutrino oscillation experiments. Therefore, if the Yukawa textures appearing in  leptogenesis can be entirely formulated by the PMNS matrix as well as the physical lepton masses, then,
 in order to generate the correct amount of Eq.~\eqref{YB}, and in particular, the positive sign of $Y_{B}^{\rm exp}$, the PMNS  structure will  be  constrained in the BAU context. In this respect, the BAU criterion may help to select the theoretical PMNS candidates from a broad class of flavor models~\cite{Altarelli:2010gt,King:2013eh} before the underlying flavor theory is found.

To apply the BAU selection criterion, we consider the mechanism presented in Ref.~\cite{Li:2021tlv}.
 \begin{figure*}[t]
 	\centering
 	\includegraphics[scale=0.58]{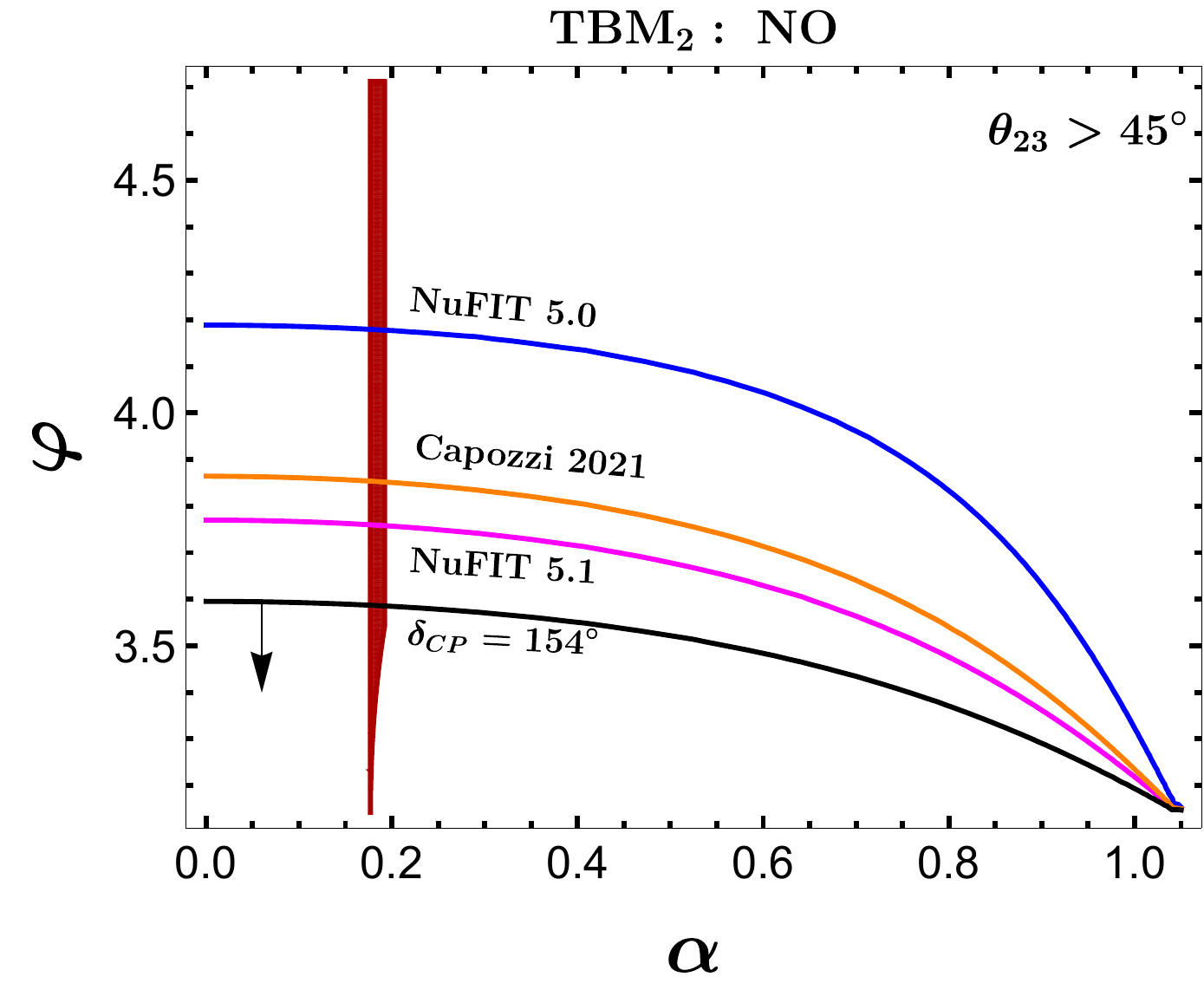}\qquad
 	\includegraphics[scale=0.58]{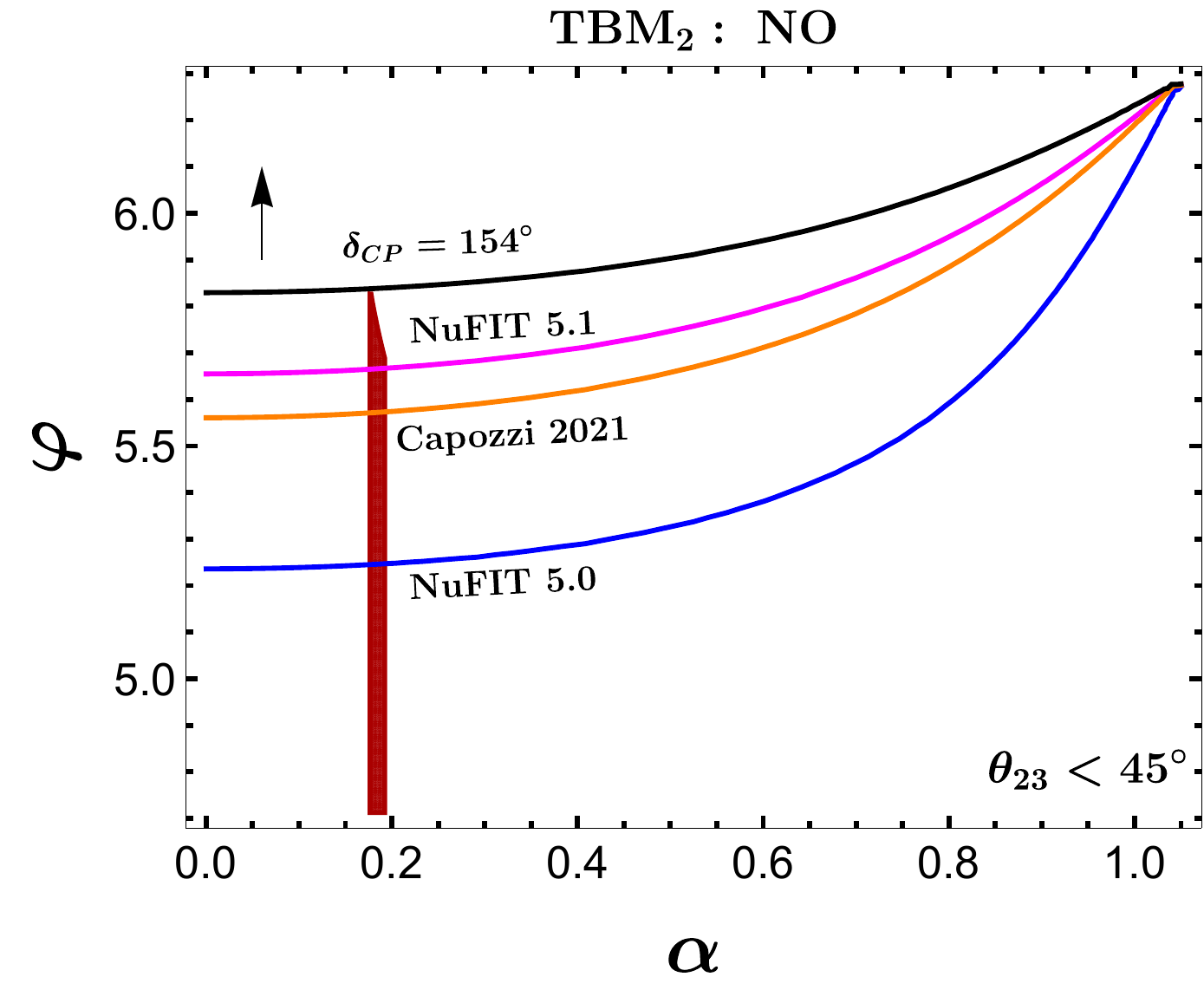}
 	\caption{\label{rR13CP23Cor} Correlation of $\delta_{CP}$ and $\theta_{23}$ under the survived $\text{TBM}_2$ pattern. The left (right) panel corresponds to the octant of $\theta_{23}>45^\circ~(\theta_{23}<45^\circ)$. The red band is allowed by  NuFIT 5.1~\cite{nufit5.1:2021,Gonzalez-Garcia:2021dve} within $3\sigma$ uncertainties, and the regions above (below) different curves are excluded by  Dirac $CP$-violating phase limits in the left (right) panel. The arrow in both panels denotes the increase of $\delta_{CP}$ towards $\pi$. Note that, similar results hold for the periodic region of $\alpha$ in the second quadrant.}
 \end{figure*}
The baryon asymmetry obtained therein has a simple dependence on the Yukawa matrix. For our purpose here, it suffices to parametrize the result as:
\begin{align}\label{YBpro}
	Y_B=\sum_{i,j,k, i\neq k}\frac{\text{Im}[Y_{\nu,i1}^* Y_{\nu,k1}(Y_\nu Y_\nu^\dagger)_{ik}] }{\vert Y_{\nu,j1}\vert^2 } \mathcal{F}_{ijk},
\end{align}
where $\mathcal{F}_{ijk}$ is a Yukawa-independent scalar function, $Y_\nu$ is  the Dirac  neutrino  Yukawa  matrix, and the index 1 denotes the fact that, the   $CP$ asymmetry is generated by a scalar decaying into the lightest neutrino (also dubbed as $\nu_1$ leptogenesis). By performing  equivalently unitary basis transformations, rather than imposing  particular structures,  on the charged-lepton and neutrino  Yukawa matrices, the leptonic  mixing can be encoded in $Y_{\nu}$, such that, \begin{align}\label{YPMNS}
Y_{\nu}\propto U_{P}\,\hat{m}_{\nu}
\end{align}
exists in the basis where the charged-lepton Yukawa matrix is diagonal. Here $\hat{m}_\nu$ denotes the diagonal Dirac neutrino mass matrix.

To see the corresponding constraint, we  apply Eqs.~\eqref{YBpro} and \eqref{YPMNS} with   the PMNS matrix now being the $\text{TBM}_{2}$  pattern.
Putting in the explicit  Yukawa-independent scalar function $\mathcal{F}_{ijk}$~\cite{Li:2021tlv}, we are led  to the following relation:
\begin{align}\label{BAU-TBM}
Y_{B}^{\text{TBM}_2}=-k\, s_{2\alpha}s_{\varphi},~
\end{align}
where $k$ is a  calculable positive parameter,  which is independent of  the free parameters $\alpha, \varphi$. From Eq.~\eqref{BAU-TBM}, we can see that, the BAU criterion  can at most  constrain the $\alpha$-$\varphi$ plane up to the sign of baryon asymmetry, since  a predicted value of $Y_{B}$ further depends on the parameter $k$.
We show in Fig.~\ref{BAUfit} the regions for $Y_{B}>0$ in the $\alpha$-$\varphi$ plane  and  compare with the survived parameter space under the constraint discussed in Sec.~\ref{CP&cons}. It can be seen clearly that, the BAU criterion  sets limits on $\varphi$, the region of which  is equivalent to requiring $\sin\delta_{CP}>0$ in Fig.~\ref{NOCPcon}.  An intriguing situation arises in the $Y_{B}>0$ region, where  two asymmetric contours coexist. We will show in the next section that, the two asymmetric contours correspond to   $\theta_{23}<45^\circ$ and $\theta_{23}>45^\circ$ octant, respectively, which are closely related to  the Dirac $CP$-violating phase.

 It is worth mentioning that, the current status of $\sin\delta_{CP}>0$ is the crucial function that renders the BAU criterion from Ref.~\cite{Li:2021tlv} available and powerful to select the unique $\text{TBM}_{2}$  pattern.
It should also be emphasized that,  if the interval for $\sin\delta_{CP}>0$ in the NO spectrum vanishes persistently  in the future, the BAU criterion from Ref.~\cite{Li:2021tlv} would become inapplicable as the  leptogenesis scenario therein  is ruled out. For $\sin\delta_{CP}<0$ in the IO spectrum, however, the logic presented here can be generalized to other Yukawa texture-independent leptogenesis, especially those support $\sin\delta_{CP}<0$, and the selection  criterion can then follow the procedure presented above.

\section{Three Unknowns  predicted by the BAU criterion}\label{Correlaions}
Currently, there are  three unknowns in neutrino oscillation experiments, namely, the neutrino mass ordering, the $CP$ violation, and the octant of atmospheric $\theta_{23}$ angle,  i.e., whether $\theta_{23}\leqslant 45^\circ$ or $\theta_{23}> 45^\circ$~\cite{Capozzi:2018ubv}. It has been shown above that, applying the BAU criterion favors an NO spectrum and $\sin\delta_{CP}>0$, and hence it can  help to  reveal the first two unknowns. In addition, since we have also  applied  the BAU criterion to obtain a quite narrow parameter space of the two parameters in NTBM patterns, it would be a   compelling  feature if the BAU criterion can further determine the third unknown.

We have observed from Fig.~\ref{BAUfit} that, there are two asymmetric contours in the $Y_B>0$ region. By considering the octant of $\theta_{23}$, we find that the small (large) contour corresponds to $\theta_{23}<45^\circ~(\theta_{23}>45^\circ)$.  It can be inferred from Fig.~\ref{NOCPcon} that, when $\delta_{CP}$ approaches $\pi$ in the NO spectrum, the allowed parameter space in the $\text{TBM}_2$ pattern would further shrink. However, the reduction speed is different in the two asymmetric contours. To visualize this, we  depict in Fig.~\ref{rR13CP23Cor} the constraints under NuFiT 5.0~\cite{Esteban:2020cvm} ($\delta_{CP}>120^\circ$, blue curve), Capozzi 2021~\cite{Capozzi:2021fjo} ($\delta_{CP}>0.77\pi$, orange curve) and NuFIT 5.1~\cite{nufit5.1:2021,Gonzalez-Garcia:2021dve} ($\delta_{CP}>144^\circ$, magenta curve), as well as  $\delta_{CP}>154^\circ$ (black curve), where  the $3\sigma$ upper limit of $\sin\delta_{CP}$ has stably decreased.  It can be seen that, when $\delta>5\pi/6$, the allowed parameter space (red band) in $\theta_{23}<45^\circ$ octant will be completely ruled out, while  a smaller region corresponding to $\theta_{23}>45^\circ$ octant survives.
 As a consequence,  the $\text{TBM}_{2}$  pattern under the BAU criterion predicts   $\theta_{23}> 45^\circ$ for $5\pi/6<\delta_{CP}<\pi$.

 The predictions of NO neutrino mass spectrum and the correlation between the  Dirac $CP$-violating phase  and the octant of $\theta_{23}$ are what we can  expect from the   $\text{TBM}_{2}$ pattern  selected from the BAU criterion, which is, however, not attainable under the constraints of mixing angles and  Dirac $CP$-violating phase   only. Besides, the predictions of three unknowns  can be readily tested in the upcoming neutrino oscillation experiments.

\section{Conclusion}
\label{cons}
We have applied two considerable constraints---the   leptonic $CP$ violation and the  BAU criterion---to the  NTBM patterns.
After imposing the constraint from Dirac $CP$-violating phase, basically   half of the existing parameter space is  ruled out. Besides, the survived regions for the rotation angle are already small even  taking the $3\sigma$ uncertainties.
For  the $3\sigma$-allowed parameter space,  we have  further applied a Yukawa texture-independent leptogenesis as a BAU criterion to select the unique NTBM pattern. It is found  that, if  the interval $\sin\delta_{CP}>0$ in the NO spectrum   keeps existing but  becomes persistently smaller,   the   BAU criterion can   phase out three of the four patterns, rendering the $\text{TBM}_2$  pattern as the only viable candidate to explain the PMNS structure.

The BAU criterion can help to uncover the three unknowns in current neutrino oscillation experiments. In addition to  favoring  the  NO spectrum of neutrino masses and  a positive $\sin\delta_{CP}$, the unique NTBM pattern under the BAU criterion  can further  predict  a clear  octant of $\theta_{23}>45^\circ$, which is correlated to the Dirac $CP$-violating phase  in the range $5\pi/6<\delta_{CP}<\pi$. All these predictions regarding the three unknowns can be fully tested in the upcoming  neutrino oscillation experiments.

The combined selection criterion from the leptonic $CP$ violation and BAU  suggested in this paper can also be  applied to other theoretical PMNS candidates, provided that the theories for flavor puzzle and the BAU generation are independent of each other.

\section*{Acknowledgements}
This work is supported by the National Natural Science Foundation of China under Grant
No.12047527 as well as by the CCNU-QLPL Innovation Fund (Grant No. QLPL2019P01) and Hubei Provincial Natural Science Foundation of China (Grant No. 2020CFB711).

\bibliographystyle{JHEP}
\bibliography{reference}

\end{document}